\documentstyle[epsfig]{mn2e}

\def\gs{\mathrel{\raise0.35ex\hbox{$\scriptstyle >$}\kern-0.6em
\lower0.40ex\hbox{{$\scriptstyle \sim$}}}}
\def\ls{\mathrel{\raise0.35ex\hbox{$\scriptstyle <$}\kern-0.6em
\lower0.40ex\hbox{{$\scriptstyle \sim$}}}}

\begin{document}

\title[Sub-mm AGN and cluster assembly] 
{The 60-$\mu$m extragalactic background radiation intensity, 
dust-enshrouded active galactic nuclei 
and the assembly of groups and clusters of galaxies} 

\author
[A.\,W. Blain and T.\,G. Phillips] 
{
A. W. Blain$^{1,2}$ and T.\,G. Phillips$^3$\\
\vspace*{1mm}\\
$^1$ Astronomy Department, Caltech 105-24, Pasadena, CA91125, USA.\\ 
$^2$ Institute of Astronomy, Madingley Road, Cambridge, CB3 0HA.\\
$^3$ Caltech Submillimeter Observatory, Caltech 320-47, Pasadena, 
CA 91125, USA.
}
\maketitle

\begin{abstract}
Submillimetre (submm)-wave observations have revealed a cosmologically 
significant population of high-redshift dust-enshrouded galaxies. The form 
of evolution inferred for this population can be reconciled easily with 
{\it COBE} FIRAS and DIRBE measurements of the cosmic background 
radiation (CBR) intensity at wavelengths longer than about 
100\,$\mu$m. At shorter wavelengths, however, the 60-$\mu$m CBR
intensity reported by Finkbeiner, Davis \& Schlegel is 
less easily accounted for.   
Lagache et al.\ have proposed that this excess CBR emission is a 
warm Galactic component, and the detection of the highest-energy 
$\gamma$-rays from blazars limits the CBR intensity at these wavelengths, 
but here we investigate possible sources of this 
excess CBR emission, assuming that it has a genuine extragalactic origin. We 
propose and test three explanations, each involving 
additional populations of luminous, evolving 
galaxies not readily detected in existing 
submm-wave surveys. First, an additional population of dust-enshrouded 
galaxies with hot dust temperatures, perhaps dust-enshrouded, 
Compton-thick active galactic nuclei (AGN) 
as suggested by recent deep {\it Chandra} surveys 
Secondly, a 
population of dusty galaxies with temperatures more typical of 
the existing submm-selected galaxies, but at relatively low redshifts. These
could plausibly be associated with the assembly of groups and clusters of 
galaxies. Thirdly, a population of low-luminosity, cool, quiescent 
spiral galaxies. Hot AGN sources and the assembly of galaxy groups 
can account for the excess 60-$\mu$m 
background. There are 
significant problems with the cluster assembly scenario, in which  
too many bright 60-$\mu$m {\it IRAS} sources are predicted. 
Spiral galaxies have the wrong spectral energy distributions (SEDs) to 
account for the excess.  
Future wide-field
far-infrared(IR) surveys at wavelengths of 70 and 250\,$\mu$m 
using the {\it SIRTF} and  
{\it Herschel} space missions will sample representative volumes of 
the distant Universe, allowing any hot population of dusty AGNs and 
forming groups to be detected. 
\end{abstract}


\section{Introduction} 

Surveys using the Submm Common-User Bolometer Array (SCUBA; Holland et 
al.\ 1999) camera at the James Clerk Maxwell Telescope (JCMT) have revealed 
a new and important population of dust-enshrouded high-redshift galaxies 
(see Smail et al.\ 2002 
and references therein). Two very 
well-studied examples of this population with redshifts are described by Ivison 
et al.\ (1998, 2000a, 2001). Distant 
dust-enshrouded galaxies have also been detected using the mm-wave MAMBO
bolometer array at 1.2\,mm (Bertoldi et al.\ 2000), and the 
{\it Infrared Space Observatory (ISO)}, both at 175\,$\mu$m using the 
ISOPHOT instrument (Kawara et al.\ 1998; Puget et al.\ 1999; 
Matsuhara et al. 2000; Linden-Vornle et al. 2000; Dole et al.\ 2001) 
and at 15\,$\mu$m using 
the ISOCAM instrument (Altieri et al.\ 1999; Elbaz et al.\ 2000).

The counts of these galaxies, and the spectrum of submm-wave cosmic 
background radiation (CBR)  
at wavelengths greater than 100\,$\mu$m (Puget et al.\ 1996; 
Fixsen et al.\ 1998; Hauser et al.\ 1998; Schlegel, Finkbeiner \& Davis 1998) 
can be readily accounted for using well-constrained models of the evolution 
of dusty galaxies at temperatures of order 40\,K (Blain et al.\ 1999c,d; 
Trentham, Blain \& Goldader 1999). 
However, the far-IR CBR intensity inferred at 
wavelengths of 140\,$\mu$m from {\it COBE}-DIRBE observations by 
Schlegel et al.\ (1998) and Hauser et al. (1998), and more recently at shorter 
wavelengths of 100 and 60\,$\mu$m by Finkbeiner, Davis \& Schlegel 
(2000), is significantly greater, by a factor of about 2, 
than that predicted by 
the models from 1999. In updated models (Blain 2001), 
based on larger datasets and with $\Omega_0=0.3$ and
$\Omega_\Lambda=0.7$, the difference between the models and Finkbeiner et al.'s 
CBR measurements are only significant at 60\,$\mu$m, as shown by 
the solid and dashed lines in Fig.\,1. Lower estimates of the CMB intensity 
at wavelengths close to the peak of the CBR spectrum 
are provided by Lagache et al.\ (1999) and Kiss et al.\ 
(2001). 

If this is a true extragalactic signal, rather than being  
associated with either a warm component of the interstellar medium (ISM) in the 
Milky Way (Lagache et al.\ 1999, 2000) or emission from a distant component of 
zodiacal dust (Finkbeiner, private communication), then this CBR data 
appears to require an additional source of far-IR, but not submm-wave, 
luminosity in the models. 

Renault et al.\ (2001) describe constraints on the CBR intensity from 
recent observations of TeV $\gamma$-ray emission from blazars.  
$\gamma$-rays are destroyed by pair production in  
interactions with far-IR photons 
that have the same numerical value of wavelength in micrometres as the 
$\gamma$-ray 
photon energy in TeV. For reasonable CBR spectra, the mean free path 
to pair production decreases strongly with increasing $\gamma$-ray energy. 
Hence, the highest-energy $\gamma$-ray photons can have 
a mean-free path through the Universe that is shorter than the 
distance to low-redshift
blazars (Aharonian 2002). 
If the 
blazar is not to have an energetically forbidden increasing power-law 
spectrum at 
high energies ($>10$\,TeV), after correcting for the intergalactic 
attenuation, then the 100-$\mu$m CBR intensity must be 
less than about 20\,nW\,m$^{-2}$\,sr$^{-1}$. However, note that the 
number of photons detected from blazars at the highest energy is small, 
and the intrinsic curvature of the blazar spectrum is unknown. A variety 
of exotic physics has also been invoked to reduce the inferred 
optical depth (see Aharonian 2002 and papers cited therein). Future 
TeV observations should allow the CBR spectrum to be mapped in detail at 
wavelengths shorter than about 100\,$\mu$m. At present it is 
worth considering models that are in mild conflict with existing results. 
If the results remain once 
more robust TeV $\gamma$-ray data is available, then the need for additional 
components of the CBR could be removed. Of course, if the 60-$\mu$m 
background excess is Galactic, then the attenuation would be negligible.  

An additional CBR component must not make a
significant contribution to the CBR intensity at wavelengths
longer than about 100\,$\mu$m. The discrete sources associated
with this excess far-IR CBR contribution also do not dominate the
counts of galaxies detected at either 1200, 850 or 450\,$\mu$m
(Bertoldi et al.\
2000; Blain et al.\ 1999b; Carilli et al. 2002; Smail et al.\ 2002), as
these results can be
accounted for in full by the existing models. However, an additional
population may
contribute to both the deep 15-$\mu$m (Altieri et al.\ 1999; Elbaz et al.\ 2000)
and 175-$\mu$m counts (Kawara et al.\ 1998; Puget et al.\ 1999). 

Note that 
the observed counts of galaxies 
from both {\it IRAS} at 60\,$\mu$m and {\it ISO} at 175\,$\mu$m 
can be 
reproduced accurately by a population of galaxies that evolves in 
luminosity as $(1+z)^{\simeq 4}$ with a dust temperature of 
about 40\,K
(see Fig.\,4 in Blain et al.\ 1999c). 
It is not possible to vary the dust temperature and form of 
evolution of a single population of galaxies to 
account for an excess 60-$\mu$m CBR
while remaining consistent with these counts. 
An additional population of 
galaxies or other sources of CBR intensity could be included, 
with a different temperature or form of evolution, but they must be  
too distant to be detected by {\it IRAS} in order to accord 
with the existing low-redshift 60-$\mu$m count data.   

Because of the thermal SED of dust-enshrouded galaxies, these conditions
imply that in order to contribute to the CBR intensity spectrum only
at the higher frequencies, either the new population has a lower mean
redshift or a greater mean dust temperature as compared with the
standard population of 40-K dusty galaxies.

Here we discuss the potential sources of this possible excess radiation. 
First, we 
consider a population of galaxies hotter than the SCUBA sources (see also 
Wilman, Fabian \& Ghandi 2000 and Trentham \& Blain 2001). These likely
correspond to optically-faint or invisible, probably dust-enshrouded, 
AGN detected in the hard X-ray band by {\it Chandra} (Fabian et al.\ 2000; 
Hornschemeier et al.\ 2000; Mushotzky et al.\ 2000). The {\it Chandra} sources 
in the background of the rich cluster A2390 are known to be hotter than the 
40-K SCUBA galaxies because their 
Rayleigh--Jeans thermal 
dust emission is not detected at 850\,$\mu$m using SCUBA (Fabian et al.\ 
2000), but their 15-$\mu$m emission on the 
short-wavelength side of their spectral peak is detected using {\it ISO} 
(Altieri et al.\ 1998): see Wilman et al.\ (2000). Dust-enshrouded AGN 
are likely to comprise at least a significant minority of the 
faint 15-$\mu$m {\it ISO} 
galaxies (Franceschini et al.\ 2002). 
Two counterexamples, cooler 
dusty galaxies with {\it Chandra} hard X-ray detections, exist 
at present (Bautz et al.\ 2000). These galaxies 
were detected by both SCUBA and {\it ISO} in the field of the cluster A370, 
with dust temperatures close to the typical 40\,K (Ivison et al.\ 1998, 
Smail et al.\ 2002). 

Secondly, we consider a population of star-forming galaxies with 40-K 
dust  
SEDs similar to those of many high-redshift 
galaxies and QSOs 
(Benford et al.\ 1999; Blain et al.\ 1999c,d; Trentham et al.\ 
1999; 
Scott et al.\ 2000), 
but which lies at a lower mean redshift. This population thus 
contributes to the CBR spectrum 
at relatively short wavelengths. A potential source of 
low-redshift activity is the conversion of the remaining cold gas in  
cluster member galaxies into stars, triggered by shocks and tidal forces 
induced 
when the galaxies are incorporated into the common dark matter halo 
of a forming group or cluster. Being 
relatively rare and short-lived events, any such less massive groups and 
more massive clusters that are forming are unlikely to 
have been observed in significant numbers in the sub-0.1-deg$^2$ 
fields that have been surveyed 
so far with SCUBA, with the exception of the central cD galaxies 
(Edge et al.\ 1999). Note that the foreground clusters in the Smail et al.\ 
(1997, 2002) SCUBA Lens Survey do not produce any significant far-IR 
emission, consistent with them being observed long after their assembly. 

Finally, we consider the CBR expected from a population of potentially 
evolving 
low-redshift spiral galaxies with lower 20-K dust temperatures (Reach et al.\ 
1995; Alton et al.\ 
1998). 

We discuss the consequences of each population for the 
spectrum of submm and far-IR CBR, 
and then consider existing and future 
observations that could discriminate between 
plausible models. In particular we consider the detection of 
hot dusty galaxies using the future wide-field far-IR and 
submm-wave surveys using the {\it SIRTF}\footnote{sirtf.caltech.edu} 
and {\it Herschel} 
(Pilbratt 1997) 
satellites. Throughout the 
paper we assume that $H_0=100h$\,km\,s$^{-1}$\,Mpc$^{-1}$ with 
$h=0.65$, $\Omega_0=0.3$ and $\Omega_\Lambda=0.7$. 

\begin{figure}
\begin{center}
\epsfig{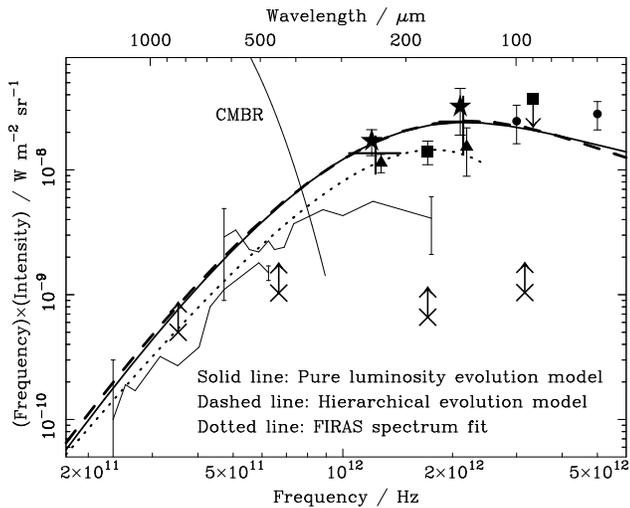}
\end{center}
\caption{The CBR intensity in the mm, submm and 
far-IR wavebands, as deduced by: Puget et al.\ (1996) -- thin solid lines 
with error bars at the ends; Fixsen et al.\ (1998) -- dotted line; Schlegel 
et al.\ (1998) -- stars; Hauser et al.\ (1998) -- thick solid crosses; Lagache 
et al.\ (1999) -- solid triangles; Kiss et al.\ (2001) -- solid squares; 
Finkbeiner et al.\ (2000) -- solid circles. 
Lagache et al.\ (1999, 2000) claim that a warm diffuse Galactic dust component 
accounts for about 50\,per cent of the isotropic DIRBE signal attributed to the 
extragalactic CBR intensity by Hauser et al.\ (1998), Schlegel et al.\ 
(1998) and Finkbeiner et al.\ (2000). The diagonal crosses represent lower 
limits to the CBR intensity inferred from source counts. From left to 
right: the 850- and 450-$\mu$m counts (Blain et al.\ 1999b; Smail et al.\ 
2002),  
and the 175- and 95-$\mu$m counts of Kawara 
et al.\ (1998) and Puget et al.\ (1999). The CBR spectrum in 
the 40-K hierarchical clustering model from Blain et al.\ (1999d) is 
represented by the solid line, while the CBR spectrum in a model 
of pure luminosity evolution (Blain et al.\ 1999c) is shown by the dashed line. 
Both curves have been updated in the light of  more 
observational data and a world model with a non-zero cosmological 
constant (see Blain 2001). 
}
\end{figure}

\section{Backgrounds from additional populations} 

\subsection{Luminous infrared galaxies with warmer dust temperatures} 

The population of luminous dust-enshrouded galaxies at moderate and high 
redshifts discussed in the context of 
modeling the SCUBA source counts and the submm-wave CBR are usually assumed 
to have dust temperatures of about 40\,K. At this temperature the observed 
low-redshift {\it IRAS} 60-$\mu$m luminosity function (Saunders et al.\ 1990) 
and the $z \ls 1$ 175-$\mu$m {\it ISO} counts can both be explained, 
requiring only a single SED and a simple form of pure luminosity evolution 
(Blain et al.\ 1999c). In an alternative hierarchical model described in 
Blain et al.\ (1999d) a temperature of about 40\,K is also required to generate 
a plausible redshift distribution for the SCUBA galaxies 
(Smail et al.\ 2002). A mean dust temperature of about 
40\,K is also consistent with 
both the SEDs of high-redshift galaxies and QSOs detected using the SCUBA, 
SHARC and MAMBO mm/submm-wave cameras
(Ivison et al.\ 1998, 2000a; Benford et al.\ 1999; McMahon et al.\ 1999; 
Carilli et al.\ 2001), the 
sample of low-redshift {\it IRAS}-selected galaxies observed by Dunne 
et al.\ (2000), Lisenfeld, Isaak \& Hills (2000) and Dunne \& Eales 
(2001) using SCUBA, and faint radio galaxies observed by 
Chapman et al.\ (2002a).  

However, an additional high-redshift subpopulation 
with a different dust temperature is not 
ruled out by the data. In fact, there is a 
significant observational bias against detecting galaxies with 
higher dust temperatures using SCUBA (Eales et al.\ 1999; Blain et al.\ 2002). 
There is so far very little information about the existence or properties of 
any hotter population. A fraction of the galaxies detected by ISOPHOT 
at a wavelength of 175\,$\mu$m 
(Puget et al. 1998) have been observed using SCUBA (Scott et al.\ 2000), but 
as their redshifts are uncertain, it was difficult to assign a reliable 
temperature; a value of $T/(1+z) \simeq 30$\,K is favoured at present. 
These ISOPHOT 
galaxies are unlikely to be at 
redshifts greater than unity. More recently, Chapman et al.\ (2002b) have 
detected two of these galaxies at $z = 0.45$ and 0.91 with dust temperatures
of only $T \simeq 30$\,K. 
One of the most interesting possibilities is 
that at least a subset of dust-enshrouded galaxies containing an AGN have 
dust temperatures hotter than the standard 
40\,K, probably reflecting a very intense 
radiation field. 
The high-redshift lensed QSOs H1413+117 (Kneib et al.\ 1998),
IRAS\,F10214+4724 (Lacy et al.\ 1998) and APM\,08279+5255 
(Lewis et al.\ 1998) have SEDs observed at wavelengths longer than 60\,$\mu$m 
that can be fitted accurately by single-temperature   
dust clouds at about 75, 80 and 110\,K 
respectively (Blain 1999a), all considerably greater than 40\,K. 
Note, however, that 
their apparent temperatures may be increased by a factor of order 
20\,per cent due to the effects of 
gravitational lensing: dust clouds on different spatial scales in different 
areas of the source can be 
magnified by different amounts. 

The lower-redshift 
{\it IRAS} galaxies P09104+4109 and F15307+3252 are also extremely 
hot, with dust temperatures greater than 100\,K 
(Deane \& Trentham 2001). At wavelengths shorter 
than 60\,$\mu$m, hotter dust components are detectable in almost all 
galaxy SEDs, but do not typically dominate the luminosity of the galaxy. 
The handful of optically faint, hard X-ray sources 
selected using {\it Chandra}, for which 850-$\mu$m SCUBA data is 
available in the same field (Bautz et al.\ 2000; Fabian et al.\ 2000;
Hornschemeier 2000; Mushotzky et al.\ 2000; Almaini et al.\ 2002) have 
submm--X-ray spectral indices consistent with dust temperatures of 
about 50\,K or greater if their redshifts are  
in the range $z \simeq 1$--2. This is all consistent 
with the ability to detect some of these X-ray-selected 
AGN in 15\,$\mu$m surveys 
using {\it ISO}-CAM (Wilman et al.\ 2000; Risaliti, Elvis \& Gilli 
2002).

We now consider the effects of adding a population of galaxies that 
are considerably hotter than 40\,K to the standard model (Blain et al.\ 
1999d).  
In the simplest case, if we add a population of hotter (80-K) 
dust-enshrouded hierarchically merging galaxies with an identical form of 
much better fit to the latest far-IR CBR intensities (Finkbeiner 
et al.\ 2000) can be obtained; see Fig.\,2. It is also quite plausible that 
a population of galaxies with a more complex distribution of 
dust temperatures that peaks at about 
80\,K could account for the observations.

The population of galaxies detected by {\it IRAS} at 60-$\mu$m extend only to 
redshift $z \simeq 0.2$, and so the addition of a hot distant population of 
galaxies would not conflict with the faint 60-$\mu$m data. Surveys
at wavelengths of 
450 and 850\,$\mu$m are more sensitive to cooler galaxies, and so 
the addition of the hotter 80-K population is not expected to conflict with the 
observed counts; however, the strength of evolution of dusty galaxies at 
$z \ls 1$ required to account for the 175-$\mu$m counts may be reduced 
slightly.

If the hotter dust-enshrouded galaxies are assumed to be 
predominantly powered by AGN accretion, and the cooler 40-K SCUBA 
galaxies are assumed to be powered by star formation, then in order 
to account for the 60-$\mu$m CBR intensity shown in Fig.\,2, 
approximately
half as much energy is generated over the history of the Universe  
by accretion as by high-mass star formation. 
As a result, assuming an accretion efficiency $\epsilon$, the ratio of 
the amount of energy released by star-formation and by accretion should be 
approximately $2 \simeq 0.007 \epsilon_* M_* / \epsilon M_{\rm BH}$, where
$\epsilon_* \simeq 0.4$ is the fraction of the material incorporated in
high-mass stars that is involved in nuclear burning,  
$M_{\rm BH}$ is the black hole mass and $M_*$ the stellar mass. If a value of 
$M_{\rm BH}/M_* \simeq 0.006$ 
(Magorrian et al.\ 1998) is Universal, then this implies that
$\epsilon \simeq 0.23$, which is 
a very plausible value. If $M_{\rm BH}/M_*$ is several 
times less  
(Ferrarese \& Merritt 2000; Gebhardt et al.\ 2000), then $\epsilon$ is 
greater by the same fraction: see also Trentham \& Blain (2001). 

A population of hotter dusty galaxies, perhaps dust-enshrouded 
AGN that are undergoing relatively efficient accretion, could thus 
account for a greater intensity of 60-$\mu$m CBR,  
without breaching other observational constraints. At wavelengths 
of 850, 450, 175 and 60\,$\mu$m  
the increase in the CBR intensity  
generated by including the additional population of hot dusty galaxies 
described above is 3.3, 5.3, 17 and 85\,per cent respectively, 
as compared with  
the conventional 40-K dusty galaxy models (Fig.\,1). 

\subsection{Luminous star-forming galaxies triggered by the assembly of 
groups and clusters} 

An alternative explanation of the excess far-IR CBR intensity is 
an additional population of low-redshift galaxies with SEDs more 
similar to those of luminous {\it IRAS} galaxies and 
high-redshift SCUBA galaxies. This population would 
boost the far-IR CBR intensity without exceeding 
the measured intensity of the submm-wave CBR. 

In hierarchical models of structure formation, the assembly of groups and 
clusters of 
galaxies is the most significant process that takes place at redshifts less 
than those at which galaxy formation takes place. The formation of clusters 
involves the merger of overdense regions, which already contain galaxies, 
into a common dark matter halo. This assembly of a new halo is likely to 
involve strong dynamical interactions of the enclosed galaxies, which are 
likely to trigger bursts of star formation activity, and perhaps consume any 
remaining cold gas in the pre-existing galaxies. The conditions in these 
member galaxies are likely to be similar to those in pairs of 
merging field galaxies, and so 
it is reasonable to assume dust temperatures similar to those in the 
40-K SCUBA galaxies for tidally- or shock-induced starbursts in forming 
clusters. 

Using a simple model of the process of cluster assembly, it is possible to 
predict the CBR spectrum expected due to this 
star formation activity. Using the Press--Schechter (Press \& Schechter 1974) 
formalism for the evolution of bound structures and the resulting merger 
rate (see Blain \& Longair 1993a,b; Blain et al.\ 1999d), the 
number of bound haloes that form at a given 
mass between $M$ and $M+{\rm d}M$, $\dot N_{\rm form}$ can be predicted 
as a function of redshift. 
\begin{equation} 
\dot N_{\rm form}=\dot N_{\rm PS} + \phi {{\dot M^*}\over{M^*}} N_{\rm PS}
\exp\left[ (1-\alpha) \left( { M \over {M^*} } \right)^\gamma \right], 
\end{equation}
where 
\begin{equation}
\dot N_{\rm PS} = \gamma { {\dot M^*} \over {M^*} } 
\left[ \left( { M \over {M^*} } \right)^\gamma - { 1 \over 2 } \right]
 N_{\rm PS}, 
\end{equation}
and 
\begin{equation}
N_{\rm PS}(M,z) = { {\bar\rho} \over {\sqrt \pi} } {\gamma \over {M^2} }
\left( { M \over {M^*} } \right)^{\gamma/2}
\exp\left[ - \left( { M \over {M^*} } \right)^\gamma \right],
\end{equation}
in which ${\bar\rho}$ is the mean density of dark matter, 
$\gamma$, $\phi$ and $\alpha$ are numerical constants, and 
$M^*(z)$ is the evolving mass of the 
typical bound halo at redshift $z$. $M^* \propto \delta(z)^{2/\gamma}$, 
where $\delta(z)$ is the amplitude of a growing density fluctuation as a 
function of redshift.  

If we assume that a bound object forming at a mass between 
$M_{\rm cl}$ and $A_{\rm cl} M_{\rm cl}$ will be recognized as a 
cluster for the first time, then the total rate of processing mass, in the 
form of dark and baryonic matter, during the formation of clusters as a 
function of redshift
\begin{equation} 
\dot \rho_{\rm cl}(z) = \int_{M_{\rm cl}}^{A_{\rm cl}M_{\rm cl}} 
\dot N_{\rm form}(M, z) M {\rm d}M. 
\end{equation}
The process of assembly of the cluster is likely to trigger large-scale star 
formation activity in the cluster member galaxies, depleting their cold gas 
content. As a result the gas depletion 
is likely to prevent any constituent galaxies 
of a cluster more massive than $A_{\rm cl} M_{\rm cl}$ from being involved 
in large-scale star formation activity at later times. $A_{\rm cl}$ is expected 
to be of order of a few, and so the integral in equation (4) can be 
approximated as 
\begin{equation} 
\dot \rho_{\rm cl} \simeq {{A_{\rm cl}^2-1}\over{2}} M_{\rm cl}^2
\dot N_{\rm form}\left[ {{A_{\rm cl}+1}\over{2}} M_{\rm cl}, z \right]. 
\end{equation} 
If we substitute 
\begin{equation} 
U = \left( { {A_{\rm cl}+1} \over 2 } { {M_{\rm cl}} \over {M^*} } 
\right)^\gamma, 
\end{equation}
then
\begin{equation}
\dot \rho_{\rm cl} \simeq 4 \bar\rho \gamma
{ {A_{\rm cl}-1} \over {A_{\rm cl}+1} } { {\dot \delta} \over \delta } 
{ {U^{1/2}} \over {\sqrt\pi} } 
\left[ \left( U - { 1 \over 2 } \right){\rm e}^{-U} + {\phi \over \gamma}  
{\rm e}^{-\alpha U} \right]. 
\end{equation} 
This expression includes the ratio of $M_{\rm cl}$ and $M^*$ through $U$. 
The ratio of these masses can be normalized, at the present epoch,  
by assuming a value for the fraction of all mass that is bound to clusters of 
galaxies at the present epoch,
\begin{equation} 
F_{\rm cl} = { { \int_{M_{\rm cl}}^\infty M N_{\rm PS} \,{\rm d}M } \over 
{ \int_0^\infty M N_{\rm PS} \,{\rm d}M } }. 
\end{equation} 
Assuming $\gamma \simeq 2/3$ (Blain et al.\ 1999d), 
if $F_{\rm cl} = 0.1$, which is reasonable for clusters, then
$M_{\rm cl}/M^*(0) \simeq 1.6$; if $F_{\rm cl} = 0.5$, which is 
reasonable for groups, then $M_{\rm cl}/M^*(0) \simeq 0.1$ (see Fig.\,3a).  
Fixing $F_{\rm cl}$ thus allows one parameter in the 
calculation to be removed. 
Note that the absolute value of $M^*(0)$ does not affect the result. 

If the fraction of cold gas in the constituent galaxies is assumed 
to be representative of the Universe as a whole at $z$,  
$\Omega_{\rm g}(z)$, and a fraction $f_{\rm g}$ of this gas is assumed to 
be converted into stars during the formation of the cluster, then the comoving 
luminosity density generated in forming clusters is 
\begin{equation} 
\epsilon_{\rm cl}(z) \simeq 0.007 c^2 
f_{\rm g} \Omega_{\rm g}(z) \dot \rho_{\rm cl}(z), 
\end{equation}   
as a function of epoch. 
The global form of evolution of the density of gas $\Omega_{\rm g}(z)$ can 
be determined by using the simple formalism discussed by Jameson (2000) and 
Longair (2000). 
In this case, 
\begin{equation} 
\Omega_{\rm g}(z) = {1 \over 2} \Omega_\infty \left\{ 1 + {\rm tanh}\left[ 
B {\rm ln}(1+z) - C \right] \right\}. 
\end{equation} 
The values of the 
two dimensionless parameters $B$ and $C$ can be  
determined by fitting the submm-wave 
CBR spectrum and the bright counts of {\it IRAS} galaxies: 
see Blain et al.\ (1999d) and Blain (2001). 
The function is an exact solution for an 
Einstein--de Sitter model, but is also an accurate representation of the 
behaviour of $\Omega_{\rm g}$ in the cosmological model assumed here
if $B=1.95$ and $C=1.60$. 

If all the energy released in star formation induced by the formation of 
a cluster is assumed to appear in the far-IR waveband, and $f_\nu$ represents 
the SED of the dusty galaxies involved, then the  
CBR intensity generated, 
\begin{equation} 
I_\nu = { 1 \over {4 \pi} } 0.007 c^2 f_{\rm g} 
\int{ {\Omega_{\rm g}(z) \dot \rho_{\rm cl}(z)} \over {1+z} }
\,{ {f_{\nu(1+z)}} \over { \int f_{\nu'} \,{\rm d}\nu' } }
{ {{\rm d}r} \over {{\rm d}z}} \, {\rm d}z,
\end{equation}
where d$r$ is the radial comoving distance element. Note that 
the value of the parameters $f_{\rm g}$ and $\Omega_\infty$ 
affect only the scaling of the predicted intensity. $A_{\rm cl}$ 
introduces a more subtle shift in the CBR spectrum. The 
values of $\phi$, $\alpha$ and $\gamma$ have only 
weak effects on the result. 

Making the simplest assumption 
of a common SED with the 40-K SCUBA galaxies, then 
the CBR spectrum predicted using equation (11) peaks at slightly too
low a frequency to describe the observations shown in Fig.\,2; however, 
if the dust
temperature is increased to 50\,K, reasonable 
agreement can be 
obtained. The results shown in Fig.\,2 are derived 
for values of $F_{\rm cl}=0.1$, 
$A_{\rm cl}=2$, 
$\gamma = 2/3$, $\phi = 1.53$, $\alpha=1.17$, $\Omega_\infty = 0.02$ and 
$f_{\rm g}=1$. 

In the context of the 
standard nucleosynthesis value of $\Omega_{\rm b} h^2 = 0.019$ (Burles \& 
Tytler 1998) and if $h=0.65$, then 
a value of $\Omega_\infty = 0.02$ corresponds to 
44\,per cent of all the baryons being available to form stars 
in the form of cold gas, which is not an unreasonable 
assumption.\footnote{A higher value of $\Omega_{\rm b}$ was suggested by an
analysis of microwave background anisotropy from the BOOMERANG experiment 
(de Bernadis et al.\ 2000); however, subsequent analysis supports a 
value of $\Omega_{\rm b}$ close to Burles \& Tytler's value (de Bernadis 
et al.\ 2002).} 
If the value of $\Omega_\infty$ is more closely linked to the mass 
of stars at the present epoch ($\Omega \simeq 0.004$ -- 9\,per cent of 
baryon density; Fukugita, Hogan \& 
Peebles 1998), then no plausible value of $A_{\rm cl}$ can account for the 
excess CBR if $F_{\rm cl}=0.1$. 

\begin{figure}
\begin{center}
\epsfig{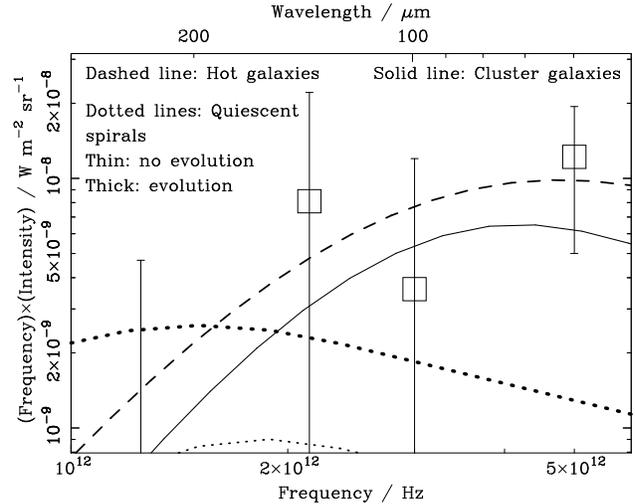}
\end{center}
\caption{CBR intensity predictions for a model of cluster assembly
(solid line), a hot population of dusty galaxies (dashed line), 
non-evolving quiescent disc galaxies (thin dotted line) and strongly 
evolving quiescent disc galaxies (thick dotted line). The form of 
evolution of the quiescent galaxies is identical to that of the standard 
40-K SCUBA galaxies. The data points are the CBR 
intensities 
reported by Finkbeiner et al.\ (2000), with the predicted contribution 
from 40-K SCUBA galaxies shown in Fig.\,1 subtracted.  
The hot galaxies are assumed to have a dust temperature of 80\,K, and a 
luminosity density evolving in the same way as, but 50\,per cent less 
intense than that of the standard SCUBA galaxy models. The 
cluster galaxies are assumed to have a temperature of 50\,K and 
$\Omega_\infty f_{\rm g} = 0.02$. The full listing 
of parameters in this model is given in Section\,2.2.   
}
\end{figure} 

The values of the parameter 
$\Omega_\infty f_{\rm g}$ (equation 
11) that is required to equal the 60-$\mu$m CBR data point in 
Fig.\,2 are shown in Fig.\,3a as a function of the mass threshold for 
cluster or group formation $M_{\rm cl}$. As the fraction of mass bound to 
structures more massive than
the hypothetical forming groups and clusters 
$F_{\rm cl}$ falls with increasing mass, 
the efficiency of star formation activity, 
parametrized by $\Omega_\infty f_{\rm g}$, must increase in order 
to generate the 
same CBR intensity. 
If the cosmic density of stars formed to generate the excess CBR is 
not to exceed the baryon density, then 
$M_{\rm cl}/M^*$ must be less than about 1.7. 
Hence, it is possible that the assembly of groups and small clusters (with 
$F_{\rm cl} \simeq 0.5$) 
could explain the far-IR CBR excess. The formation of rarer massive 
clusters cannot account for the excess, because too great a  
baryon mass would have to be involved. 

In the `standard' model of dusty galaxy evolution, the density of 
stars generated exceeds the value of Fukugita et al.\
(1998) by a factor 2--3 (see fig.\,6 in Blain et al.\ 1999d). 
This discrepancy can be corrected relatively easily by requiring 
the initial mass function (IMF) to have a lower limit of 
1--3\,M$_\odot$. In the case of clusters with $F_{\rm cl} = 0.1$ 
$(M_{\rm cl} / 
M^* \simeq 1.6)$, 
however, the 
discrepancy between the 
mass processed into stars in order to produce the observed CBR 
intensity exceeds the 
Fukugita et al. measurement by a factor of about 10. 
No changes in the form of the IMF can reconcile these values. 

Any excess star-formation activity induced by group and cluster assembly is 
likely to be 
rather short lived, and so not to introduce any significant variation in the 
colours of the ensemble of cluster member galaxies. It is thus unlikely 
that observations of cluster and group member galaxies made 
several Gyr after their
formation epoch would be able to detect any clear signs of this final burst of 
activity (Verdes-Montenegro et al.\ 1998). 
Observations that might probe the evolution of cluster member 
galaxies at moderate redshifts include searches for signs of ongoing 
star formation activity, via either unobscured 
starlight (Poggianti et al.\ 1999), radio emission from supernova remnants 
(Dwarakanath \& Owen 1999) or mid-IR emission from heated dust (Fadda et al.\ 
2000). 

\begin{figure*}
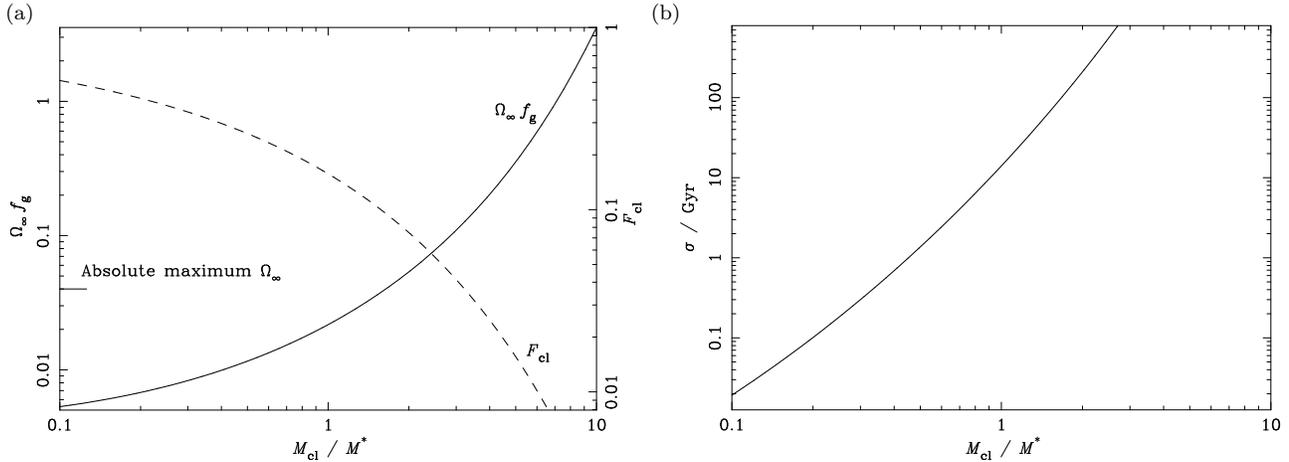

\begin{minipage}{170mm}
(a) \hskip 81mm (b)
\begin{center}
\vskip -5mm
\epsfig{file=fig3a.ps,width=5.8cm,angle=-90} \hskip 4mm
\epsfig{file=fig3b.ps,width=5.8cm,angle=-90}
\end{center}
\caption{(a) Solid line: the value of the parameter $\Omega_\infty f_{\rm g}$ 
(equation 11; 
left-hand ordinate axis) 
in the cluster/group assembly model that is 
required to produce an excess 60-$\mu$m CBR intensity equal to that 
shown by the 
data point in Fig.\,2. The value of $\Omega_\infty = \Omega_{\rm b}$ for 
$f_{\rm g}=1$ is shown as an absolute maximum value of this parameter. 
The value of the parameter $F_{\rm cl}$ (equation 8; right-hand ordinate 
axis) is also shown. As $F_{\rm cl}$ declines, $\Omega_\infty f_{\rm g}$ 
must increase in order to generate the same CBR intensity.
(b) The value of the parameter $\sigma$ (equation 16) that is required for a 
model of cluster formation that reproduces the excess 60-$\mu$m 
CBR intensity to equal the bright 60-$\mu$m {\it IRAS} count.
Values of $\sigma$ below the line are inconsistent with the count observations. 
Realistic models must have values of $\sigma$ much less than the Hubble time. 
}
\end{minipage} 
\end{figure*} 

\subsection{Low-redshift quiescent disc galaxies} 

The population of low-redshift disc galaxies similar to the Milky Way, 
with dust temperatures of about 20\,K (Reach et al.\ 1995; Alton et al.\ 1998), 
can be associated with either the low-luminosity end of 
the empirical 60-$\mu$m {\it IRAS} galaxy luminosity function 
(Saunders et al.\ 1990; Soifer \& Neugebauer 1991), or with the mass function 
of non-merging galaxies derived in an hierarchical clustering model of 
galaxy evolution, assuming a certain mass-to-light ratio (see Fig.\,11a of 
Blain et al.\ 
1999d). The observational constraints on such a population are discussed 
further by Barnard \& Blain (2002); note that its properties are 
constrained by the counts of IRAS galaxies at 60\,$\mu$m (Bertin, 
Dennefeld \& Moshir 1998), and the bright counts at 175-$\mu$m  
reported by Stickel et al.\ (1998).   

In Fig.\,2 we show the CBR intensity predicted by a 
suitably constrained population of quiescent galaxies, drawn from 
a Press--Schechter mass function of galaxies, assuming a dust 
temperature of 20\,K and a mass-to-light ratio during ongoing quiescent 
star-formation activity that is 1\,per cent of the rate during a merger -- 
values 
that are consistent with counts of bright {\it IRAS} and {\it ISO} galaxies. 
Regardless of whether the 
population of quiescent galaxies undergoes either 
no evolution or strong evolution 
of a form that matches the evolution of the luminosity density of merging 
galaxies, the CBR spectrum attributable to this quiescent 
population is quite different to the observed excess (Finkbeiner et al.\ 2000), 
both in terms of its SED and its absolute intensity. Hence, no 
plausible form of evolution of a population of 
dusty spiral galaxies can account for the excess far-IR CBR
intensity. 

If the properties of the population of quiescent, non-merging galaxies were 
modified to generate a CBR 
spectrum similar to that observed, then evolution of the same form as that of 
the merging galaxies, hotter dust temperatures of about 40\,K, and an increase 
in the intensity of star-formation activity by a factor of about 5, are all 
required. In this case, the properties of the `quiescent' galaxies become 
indistinguishable from those of the merging dusty galaxies included in 
existing models, and so they cease to make up a distinct, 
additional population. 
A population of disc galaxies with enhanced 
star formation rates due to their infall into clusters could produce a suitable 
CBR spectrum, but the number of galaxies falling into clusters 
(Meusinger, Brunzendorf \& Krieg 2000) would be far 
too small to account for the factor of two increase in the CBR 
intensity required to account for the 60-$\mu$m observations of 
Finkbeiner et al.

\subsection{Permitted models} 

A co-ordinated burst of star-formation activity induced by the assembly 
of small clusters and groups of galaxies, and a hot population of 
dust-enshrouded 
merging 
galaxies both appear to be able to reproduce the possible excess in the 
measured spectrum of the 
far-IR CBR. 

No current observations rule out the hot galaxies model. The CBR intensity 
in this model is generated at $z \simeq 1$ by galaxies 
with luminosities typically of several $10^{12}$\,L$_\odot$, and so these 
galaxies will not be found in either 60-$\mu$m {\it IRAS} counts (Bertin 
et al.\ 1997) or 
bright 175-$\mu$m 
{\it ISO} counts (Stickel et al.\ 1998). However, they would contribute 
to the faint 15-$\mu$m {\it ISO} counts and the counts of 
faint hard X-ray {\it Chandra} galaxies: the majority of both these 
populations appear to galaxies at 
$z \simeq 1$. The acquisition of much larger 
samples of these objects, using {\it Chandra}, {\it XMM--Newton} and 
{\it SIRTF} will be very valuable for testing this hypothesis. 

Conversely, a ready test is available for the idea that co-ordinated star 
formation activity that takes place at $z < 1$
in assembling groups and clusters generates the CBR excess. 
The amount of activity required to 
reproduce the far-IR CBR can be determined, and by assuming 
a time-scale for the duration of the activity, a count of far-IR luminous 
clusters in the process of formation can be derived. 
The measured bright counts of galaxies at 60 and 175\,$\mu$m can be 
predicted. As these counts are already explained by 
the evolving field galaxy models,   
the group/cluster assembly scenario must not lead to a large additional 
contribution.

\subsection{Far-IR counts of assembling clusters} 

From equation (1), we know the formation rate of clusters. The 
comoving space density of clusters forming at masses between $M_{\rm cl}$ 
and $A_{\rm cl} M_{\rm cl}$ is
\begin{equation} 
n(z) = \sigma \int_{M_{\rm cl}}^{A_{\rm cl}M_{\rm cl}}
\dot N_{\rm form}(M,z) \, {\rm d}M,  
\end{equation}
where $\sigma$ is assumed to be a constant duration of the luminous phase. 
As $A_{\rm cl}$ cannot be too large, this integral can be approximated by, 
\begin{eqnarray}
\lefteqn{\nonumber
n(z) \simeq { {8 \sigma \bar\rho \gamma} \over  {\sqrt\pi} } 
{ {\dot \delta} \over \delta } {{A_{\rm cl}-1} \over {(A_{\rm cl}+1)^2} } 
\times } \\ 
& & \>\>\>\>\>\>\>\>\>\>\>\>\>\>\>\>\>\>\>\>\>\>\>\>\>\>\>\>\>\>
\left[ \left(U - {1 \over 2}\right) e^{-U} + {\phi \over \gamma} e^{-\alpha U}
\right] { {U^{1/2}} \over {M_{\rm cl}} }, 
\end{eqnarray} 
where $U$ was defined in equation (6). 
This expression can be integrated over comoving 
volume to yield the count of such objects on the sky brighter than a flux 
density $S_\nu$ at frequency $\nu$ per unit solid angle, 
\begin{equation} 
N(>S_\nu) = \int_0^{r_{\rm max}(S_\nu)} n[z(r)]  r^2 \, {\rm d}r, 
\end{equation}
where,  
\begin{equation} 
r_{\rm max} \simeq \left[ { {0.007 c^2} \over {\sigma}} f_{\rm g} 
{ {A_{\rm cl}+1} \over {4 \pi S_\nu } } { {M_{\rm cl}} \over 2 } 
\Omega_{\rm g}(0) 
{ {f_\nu} \over {\int f_\nu' \, {\rm d}\nu'}} \right]^{1/2}, 
\end{equation} 
the maximum distance to which a star-forming group or cluster can
be seen. By evaluating equation (14) at  
$z \simeq 0$, which is a reasonable redshift 
to assume for the brightest counts, 
\begin{eqnarray}
\lefteqn{\nonumber
N(>S_\nu) = { {\bar\rho \gamma c^3} \over {12 \sqrt{2} \pi^2}} 
{{A_{\rm cl}-1} \over {\sqrt{A_{\rm cl}+1}} } { {\dot \delta} \over \delta }
\left[ { { M_{\rm cl} U } \over \sigma } \right]^{1/2} S_\nu^{-3/2} \times } \\
& & 
\left[ { {0.007 f_{\rm g} \Omega_{\rm g}(0) {f_\nu} } \over {\int f_{\nu'} \, 
{\rm d}\nu'}} \right]^{3/2}
\left[ \left(U - {1 \over 2}\right) e^{-U} + {\phi \over \gamma} e^{-\alpha U} 
\right]. 
\end{eqnarray}
The only parameter not constrained explicitly from the fit to the excess CBR 
spectrum shown in Fig.\,2 is $M_{\rm cl}$, for any assumed dust 
temperature and thus SED. The effect of changing the value of $M_{\rm cl}$ 
on the predicted count is very great: the value of $\sigma$ required to 
generate both the excess 60-$\mu$m CBR shown in Fig.\,2 and 
the observed 60-$\mu$m count $N(>S_\nu)$ at the bright flux density 
$S_\nu = 5$\,Jy is 
shown in Fig.\,3b. 
Note that the value of $M_{\rm cl} \simeq 3.6 \times 10^{12}$\,M$_\odot$ 
based on the Tully--Fisher normalized value of $M^*$ on galactic 
scales (Blain, M\"oller \& Maller 1999). 

The observed 175-$\mu$m count, $N_{175} \simeq 1800$\,sr$^{-1}$ 
brighter than $S_{175} = 2.2$\,Jy  
(Stickel et al.\ 1999), and the 60-$\mu$m count,
$N_{60} \simeq 
60$\,sr$^{-1}$ 
brighter 
than $S_{\rm 60}=5$\,Jy (Bertin et al.\ 1999), can be compared with 
the predicted ratio from equation (16),  
\begin{equation}
{ {N_{175}(S_{175})} \over { N_{60}(S_{60})} } = 
\left( { {f_{60}} \over {f_{175}} } \right)^{-3/2}
\left( { {S_{60}} \over {S_{175}} } \right)^{3/2}.  
\end{equation}  
Taking a reasonable ratio of the SED values $f_{60}$ and 
$f_{175}$ ($f_{60}/f_{175} \simeq 4.3$ for 50-K dust), at the flux 
limits $S_{175}$ and $S_{60}$ listed above, the 
predicted count ratio  
$N_{175}/N_{60} = 0.38$. Hence, the 
most significant constraint is imposed by the requirement to agree 
with the 60-$\mu$m 
count. 

In order not to exceed the 60-$\mu$m count, while generating the 
excess 60-$\mu$m CBR value shown in Fig.\,2, a 
value of $\sigma=80$\,Gyr is required for 
$M_{\rm cl}/M^* = 1.6$ ($F_{\rm cl} = 0.1$), while for  
$M_{\rm cl}/M^*=0.1$ ($F_{\rm cl} = 0.5$), $\sigma \simeq 19$\,Myr is 
required. Note that 
the count 
depends on $\sigma^{-1/2}$, and so relatively large changes in 
the value of $\sigma$ are required to modify the count significantly. 
Unless $\sigma$ is significantly greater than the value shown in Fig.\,3b, 
the fraction of bright 60-$\mu$m {\it IRAS} counts due to assembling 
groups and 
clusters 
will be unacceptably large. Hence, only for groups with 
values of $M_{\rm cl}/M^* \ls 0.2$ is 
it possible to generate excess 60-$\mu$m CBR without violating the 
60-$\mu$m count constraint. The large value of $\sigma$ required in the 
cluster model renders it capable of accounting for only an 
very small fraction of the CBR excess. 

The constraint on $\sigma$ can be relaxed slightly if the  
individual galaxies in the cluster or group undergo star-formation 
activity that is not coincident in time. The count slope is $-3/2$ 
at bright flux densities, and so splitting the activity into $N$ 
components should lead to a reduction in the count by a factor of 
$\sqrt N$, and 
thus to a reduction in the limiting value of $\sigma$ by a factor of $N$. 
However, even if $N \simeq 100$, $\sigma$ remains at 0.8\,Gyr for 
clusters with $F_{\rm cl} \simeq 0.1$. When the requirement for 
non-simultaneous bursts in each galaxy is multiplied in, the total 
duration of star formation activity in the cluster remains 
comparable to or greater than the Hubble time, and so unlike the assembly of 
groups, the assembly of clusters is  
not a plausible mechanism 
for explaining the 60-$\mu$m CBR spectrum. The amount of dust-enshrouded 
star-formation activity required to generate the additional 60-$\mu$m 
CBR spectrum would lead to a large number of mid-IR-luminous  
clusters on the sky in this model, 
which were not observed by {\it IRAS}. 

\section{Observational confirmation of a 
population of hot dusty galaxies or forming groups} 

From the discussions above, we suspect that a population of 
dust-enshrouded galaxies at cosmological distances with dust 
temperatures that are considerably higher than inferred for most nearby 
{\it IRAS} galaxies and high-redshift SCUBA galaxies could be 
responsible for 
the excess far-IR CBR intensity reported by Finkbeiner et al.\ 
(2000). A population of assembling galaxy groups with masses a few times less 
than $M^*$ could also be a plausible source of the excess far-IR CBR, if 
star formation was induced in the gas-rich member galaxies during the 
interactions
associated with the assembly of the group. 
Any contribution from 
our alternative proposals -- cooler evolving dusty 
galaxies with SEDs such as spiral galaxies, 
or star formation induced in the assembly of 
rich clusters --  
are unlikely to be significant. 

In order to confirm the presence of a significant population of hot 
dusty galaxies, 
sensitive large-area surveys are required at short 
submm and far-IR wavelengths. Deep images are required in order 
for distant galaxies 
to be detected, and images of the survey fields in several wavebands 
spanning the mm to mid-IR wavebands are required in order to place 
constraints on the redshifts and temperatures of the detected objects (see 
Blain 1999c and Blain et al.\ 2002). 
A dataset that might be sufficiently deep for 
conducting this research has been obtained using {\it ISO} (Oliver et al.\ 
2000). However, much larger images, with much better 
photometric accuracy, resolution and spectral coverage, will be 
produced by the {\it SIRTF} telescope in 2003, the SOFIA airborne 
observatory after 2004/5 and ultimately by the 3.5-m 
aperture {\it Herschel} telescope at the end of the decade. 

Note that the thermal nature of the emission from these objects means that 
even the most accurately calibrated multiwavelength observations across 
the submm, far- and mid-IR wavebands can only provide information 
about the redshifted dust temperature $T/(1+z)$, and not the temperature 
directly. A dust temperature can only be derived once either a photometric 
redshift, determined from accurate multicolour photometry in the 
optical/near-IR wavebands, or a spectroscopic redshift is obtained. 
This temperature--redshift degeneracy is also expected to affect the 
radio--far-IR flux ratio in the detected galaxies (Carilli \& Yun 1999; 
Blain 1999b) if $T < 60$\,K. 

In order to test the idea of forming groups being responsible for excess 
60-$\mu$m CBR intensity, 
the spatial distribution of luminous dusty galaxies within a representative 
cosmological volume must be determined. This can be derived once 
areas in excess of order 1\,deg$^2$ have been surveyed out to a redshift 
$z \gs 1$. 
Observations of fields of similar size at longer wavelengths 
using future wide-field 
mm and submm-wave instruments, including the 1.1/1.4/2.1-mm BOLOCAM 
(Glenn et al.\ 1998), 450/850-$\mu$m 
SCUBA-II (Holland et al.\ 2000) and 350-$\mu$m SHARC-II (Moseley, Dowell 
\& Phillips 2000) 
cameras will provide high-resolution images of these 
fields at longer wavelengths, and thus accurate SEDs for the detected 
sources. 
The analysis of 
clustering in, and the determination of  
optical counterparts to and redshifts for, galaxies detected 
in unbiased deep far-IR surveys covering several square degrees 
using 
{\it SIRTF} and {\it Herschel} will be crucial for testing 
whether a significant fraction of 
faint 60-$\mu$m sources are associated with forming groups, and 
so this 
explanation for an excess CBR could be correct. 

At present the clustering properties of submm and far-IR selected 
galaxies have not been investigated 
using large statistical samples, owing 
to the small number of detected objects, 
although a start has been made using analyses of
background fluctuations and correlation functions in observations using 
SCUBA (Peacock et al.\ 2000; 
Almaini et al.\ 2002; Scott et al.\ 2002) 
and fluctuations using {\it ISO} (Lagache \& Puget 2000; Kiss et al.\ 
2001). 
No region containing highly clustered 
sources, as might correspond to groups or clusters in the process of 
formation has been detected in an unbiased survey in these wavebands, 
as the surface 
density of such objects is expected to be small, and an insufficient area 
of sky has so 
far been surveyed to the appropriate depths to detect one. 
Note that there are 
signs from observations in the fields of powerful high-redshift 
radio galaxies, which may be tracers of high peaks in the density field of the 
early Universe (Ivison et al.\ 2000b), that very significant overdensities 
of dust-enshrouded galaxies do exist at high redshifts. 

\section{Conclusions} 

If recent reports of additional extragalactic background radiation at an 
observed wavelength of 60\,$\mu$m are correct, then an extra 
population of dust-enshrouded galaxies could be required in models of 
galaxy formation in order to account for them. This is in addition to the 
relatively cool 40-K dusty galaxies that can account for the measured submm 
and far-IR counts and the mm- and submm-wave CBR
intensity. The most plausible sources of the excess emission are 
likely to be either a population of hotter dust-enshrouded galaxies, heated 
predominantly by AGN, or a population of 
star-forming galaxies associated with the 
assembly of galaxy groups. 
It is very unlikely that dust emission from either 
moderate redshift spiral galaxies or dust-enshrouded star formation in 
assembling rich clusters of galaxies can account for the excess 
CBR intensity. 
Follow-up observations of faint X-ray sources 
detected by {\it Chandra} and {\it XMM-Newton} using mid-/far-IR 
sensitive telescopes, such as {\it SIRTF} and {\it Herschel}, and 
deep wide-area 
surveys of representative cosmological volumes using the same telescopes,  
should allow these suggestions to be 
be confirmed or refuted. 

\section*{Acknowledgements} 

We thank Doug Finkbeiner, for discussing his CBR intensity 
estimates prior to publication, Ian Smail for an insightful referee's 
report, and Kate Quirk and Neal Trentham for helpful 
comments on the manuscript. In Cambridge, AWB was supported by the 
Raymond and Beverly Sackler Foundation as part of the Foundation's 
Deep Sky Initiative Programme at the IoA. TGP acknowledges the support 
of NSF through grant \#AST 9980846 to the Caltech Submillimeter Observatory 
(CSO).

\end{document}